# Mechanical Properties and Fracture Phenomenon of Defected Monolayer Indium Selenide: A Molecular Dynamics Study

Md. Faiyaz Jamil,[a] Md. Sagir A.M. Jony,[a] Tanmay Sarkar Akash,[a] Rafsan A.S.I. Subad,[a] and Md Mahbubul Islam*[b]

[a]Department of Mechanical Engineering, Bangladesh University of Engineering and Technology, Dhaka-1000, Bangladesh

[b]Department of Mechanical Engineering, Wayne State University, 5050 Anthony Wayne Drive, Detroit, MI- 48202, USA

*Corresponding Author Tel.: 313-577-3885; E-mail address: mahbub.islam@wayne.edu

**ABSTRACT**

In this study, we report the mechanical properties and fracture mechanism of pre-cracked and defected InSe nanosheet samples using molecular dynamics (MD) simulations. We noticed that the failure of pre-cracked and defected InSe nanosheet is governed by brittle type fracture. Armchair directional bonds exhibit a greater resistance for crack propagation relative to the zigzag directional ones. Thus, fracture strength of the pre-cracked sheet is slightly higher for zigzag directional loading than that for armchair. We evaluated the limitation of the applicability of Griffith's criterion for single layer (SL) InSe sheet for nano-cracks as the brittle failure of Griffith prediction demonstrates significant differences with the MD fracture strength. We inspected the effect of temperature on the mechanical properties of the pre-cracked samples of SLInSe. We also discussed the fracture mechanism of both defected and pre-cracked structure at length.

## 1. INTRODUCTION

Graphene, the first 2D material fabricated in 2004 has engendered a new era in material science [1]. Graphene demonstrates different extra-ordinary properties like atypical quantum Hall effect [2], outstanding crystal quality[3], higher electron mobility [4], superconductivity [5], etc. which makes it eligible for a large area of application. These outstanding properties of graphene made researchers and scientists much more enthusiastic to search for and fabricate other 2D materials as well. As a result, a bunch of graphene analogues such as Silicene [6], Germanene [7], Hexagonal boron nitride [8], etc and a numerous number of 2D material groups i.e. Transition Metal Dichalcogenides (TMDCs) [9], 2D pnictogens [10], etc were discovered over the past few years. Recently, a novel TMD, Indium Selenide (InSe) has been exfoliated from bulk InSe to atomic thick layers [11][12].

Exfoliated InSe is a van-der-Waals layered III-VI metal chalcogenide with an atomically smooth surface and high thermal stability up to 660°C [13]. This is also stable under ambient conditions as no decomposition in air is observed [14]. According to density functional theory(DFT), bulk InSe has a band gap of about 0.41 eV [15] which can be transformed into an indirect semiconductor when exfoliated to monolayer or a few layers, accompanied by a large increase in the band gap [16] which overcomes the disadvantage of graphene whose bandgap is zero [17]. Its proven high mobility [13][18] makes it eligible for application of Field effect transistors. Moreover it shows highly promising optoelectronics properties like photo sensor[19], high responsivity photodetector (450–785nm),[20][21][22] etc. Besides, first-principles calculation indicates that piezoelectric coefficient of InSe sheet is of same magnitude compared to other 2D materials, which open the possibility for this monochalcogenide semiconductor to integrate elcetromechanical and optical sensors on the same material platform [23].

These fascinating properties have motivated researchers to continue more investigation on this material. As a result, many studies have been performed to understand its structural stability [14], electronic structure [24][16], photoluminescence [25], and surface photovoltaic effect [25] through the methods such as magnetic field [26], controllable oxidation [27], high pressure [28] and so on. As uniaxial strain can tune the optical and electronic properties of monolayer InSe [29], recently, an MD study has been carried out by *Chang et al.* [30] to investigate the influence of temperature and strain rate on the mechanical properties and fracture process of monolayer InSe. However, from the applications of monolayer InSe, it is evident that frequent exposure to higher temperatures and unfriendly chemical environment make it susceptible to the evolution of defects [31][32]. Moreover, in real-life manufacturing process, there might be some inherent structural defects. These defects are an integral part of the structure and are inevitable. Sometimes these defects might be introduced to the structure willingly to control the electrical, mechanical, and optoelectrical properties [33,34]. Thus, it is essential to inspect the impact of defect, and crack in the InSe nano-sheet to find out their mechanical responses. To our best knowledge, there is hardly any MD study regarding the defected, and pre-cracked structure of SLInSe.

In this work, we studied the mechanical properties of SLInSe with existing defects, and cracks. Here, we have considered point defects of various concentrations and observed how these affects the stress-strain relationship. We incorporated pre-cracks of length from ~2.5 nm to ~5.4 nm and stress-strain relationship is obtained. Calculated fracture stress is then compared with the Griffith's theory. We have also investigated the temperature dependency of fracture strength of SLInSe with fixed initial defects. The simulations are carried out at different temperatures ranging from 1K to 600K. The fracture mechanism of the defected structures is investigated in detail.

## 2. METHODOLOGY

We generated 30nm x 30nm SLInSe sheets using MATLAB script [35]. The total number of atoms is around 25000. The covalent bond between In and Se atoms gives rise of a honeycomb array in xy plane. The three important parameters- lattice constant, the bond lengths for In-Se and In-In are 4.09 Å, 2.69 Å and 2.82 Å respectively [36]. The effective thickness of the sheet is 8.32 Å [37]. We used Atomsk software [38] to create single vacancy, bi vacancy, and varied defects concentration by arbitrarily removing atoms. With the previous studies, it is evident that in order to avoid size effects of finite dimension, the minimum dimension of nano-sheet must be more than ten times the half crack length [39][40]. Here, five cracks of different lengths (approximately ~ 2.5 nm, ~ 3.3 nm, ~ 4.1 nm, ~ 4.8 nm and ~ 5.4 nm) were placed by deleting atoms in the center of the nanosheet both in armchair and zigzag direction with the help of OVITO [41] software to study the effect of crack length on the fracture strength of InSe nanosheet.

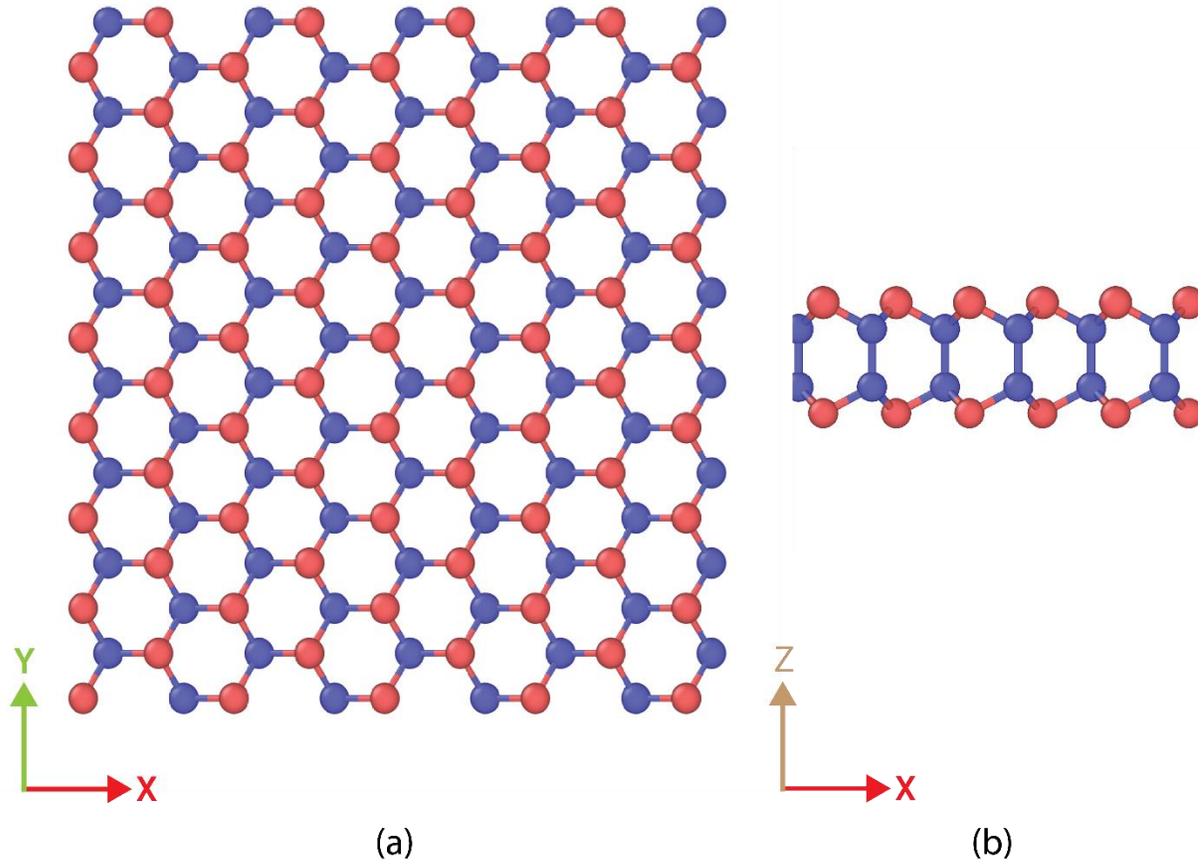

Fig. 1. SLInSe (a) Top view (b) side view. Red, and blue colors represent Se, and In atoms accordingly. Armchair direction is aligned with the x-axis and the zigzag direction with the y-axis

We employed Molecular Dynamics (MD) simulation to study the mechanical properties and the fracture mechanism of SLInSe. All simulations were performed using LAMMPS [42] software package. Periodic boundary condition is applied in the planar direction (x and y) which eliminates the finite-length effect; z-direction is kept non-periodic. Time step is selected to be 1 fs. Initially, energy minimization of the structure is ensured using the conjugate gradient (CG) algorithm. After the energy minimization, the structure is relaxed using isothermal-isobaric (NPT) ensemble for .1 ns at a specific temperature by using the Nosé–Hoover algorithm. Afterwards canonical (NVT) ensemble is used for .1 ns at the same temperature. The damping constants selected are .5 ps for pressure in both x and y direction and 10 ps for temperature. Here, we select the constant axial strain rate $10^9 \, s^{-1}$. Apparently, this strain rate is much higher compared to the experimental studies, but many contemporary studies use the strain rate of similar range [43][44]. With constant strain rate the simulation box is deformed and average stress is quantified over the structure. Here, stress is calculated using the measurement of virial stress. The relationship used to calculate the virial stress is [45]:

$$\boldsymbol{\sigma}_{virial} = \frac{1}{\Omega} \sum_i \left( -m_i \dot{u}_i \otimes \dot{u}_i + \frac{1}{2} \sum_{j \neq i} r_{ij} \otimes f_{ij} \right) \qquad (1)$$

Here, $\Omega$ is the total volume occupied by the atoms of the system, $m_i$ represents the mass of atom $i$, $\dot{u}_i$ is the time derivative representing the displacement of an atom with respect to a reference position, $r_{ij}$ represents the position vector of an atom, $f_{ij}$ is the interatomic force applied on atom i by atom j, and $\otimes$ symbol represents the cross multiplication. As the stress is calculated without considering the actual thickness of the structure, so for simplicity N/m is used instead of Pa unit.

Engineering strain can be calculated using the following formula:

$$\varepsilon = \frac{l - l_0}{l_0} \tag{2}$$

where, $l_0$ is the initial length and $l$ is the instantaneous length of the simulation box.

Atomic interactions in this study were described by the Stillinger-Weber (SW) potential [46]. There are a two-body term and a three body term in SW potential expressing the bond stretching and bond breaking accordingly. The mathematical expressions are as follows:

$$\Phi = \sum_{i<j} V_2 + \sum_{i>j<k} V_3 \tag{3}$$

$$V_2 = A e^{\left[\frac{\rho}{r - r_{max}}\right]} \left(\frac{B}{r^4} - 1\right), \tag{4}$$

$$V_3 = K \theta e^{\left[\frac{r_1}{r_{ij} - r_{max\ ij}} + \frac{r_2}{r_{ik} - r_{max\ ik}}\right]} (\cos q - \cos q_0)^2 \tag{5}$$

Here, $V_2$ and $V_3$ denote the two-body bond stretching and angle bending terms respectively. The terms $r_{max}$, $r_{max\ ij}$, $r_{max\ ik}$ mention cutoffs. Angle between two bonds at equilibrium configuration is indicated by $\theta_0$. A and K are energy related parameters which are established by Valance Force Field (VFF) model. B, $\rho$, $\rho 1$, and $\rho 2$ are other parameters that are fitted coefficients. Their value are obtained from ref. [46].

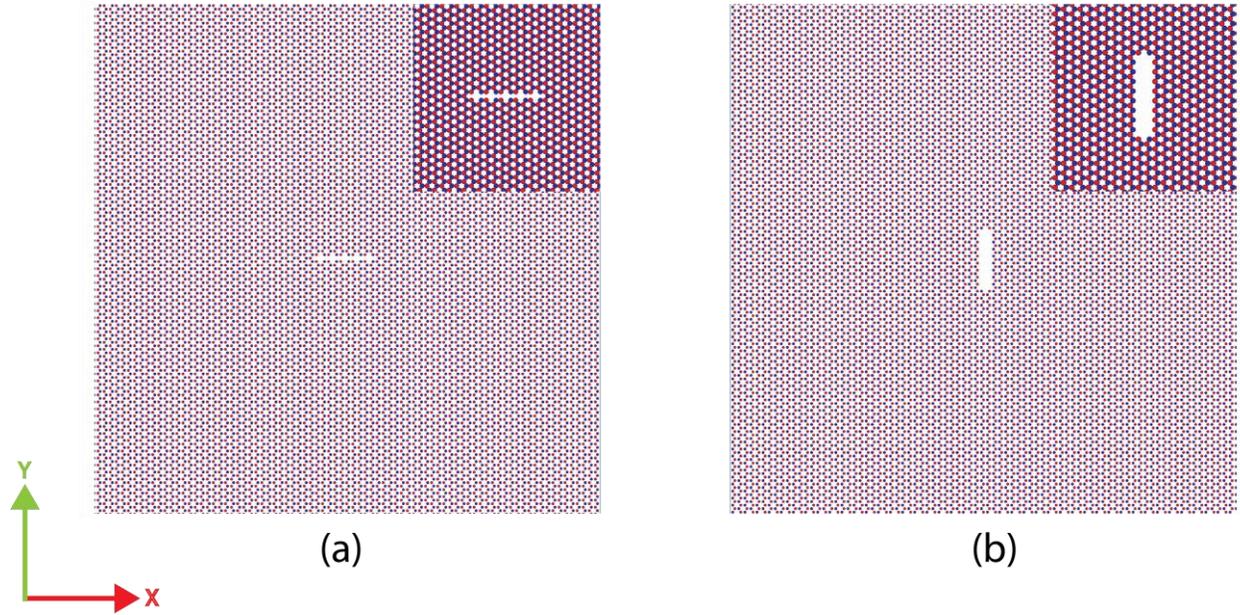

Fig. 2. Atomic structure of SLInSe (a) Crack in the armchair direction (b) Crack in the zigzag direction. Blue color represents the In atoms and red color represents the Se atoms.

## 3. METHOD VALIDATION

To validate the approach used in this study, uniaxial tension is applied both in the armchair and zigzag direction of the pristine sheet of SLInSe and their stress-strain relationship has been depicted in the Fig. 3. The obtained results are compared with previous literature [30] in Table1. The comparison indicates that the calculated results in this study are in well agreement with the existing literature.

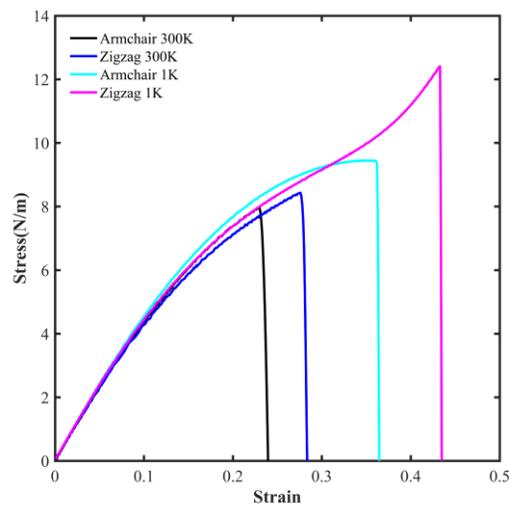

Fig 3. Stress-strain relationship of pristine SLInSe

Table 1 Validation of this study with the existing one

| Mechanical Properties | Current study | | Previous study[30] | |
|---|---|---|---|---|
| | Armchair loading | Zigzag loading | Armchair loading | Zigzag loading |
| **Young's Modulus (N/m)** | ~44.13 | ~41.3 | 45.7 | 44.8 |
| **Ultimate Tensile Stress (N/m)** | 9.45 (1K)<br>7.97 (300K) | 12.41 (1K)<br>8.43 (300K) | ~9.8 (1K)<br>8.6 (300K) | ~12 (1K)<br>8.3 (300K) |
| **Fracture Strain (%)** | 36.2 (1K)<br>23.3 (300K) | 43.3 (1K)<br>27.7 (300K) | ~36 (1K)<br>24.45 (300K) | ~44 (1K)<br>26.86 (300K) |

## 4. RESULTS AND CALCULATIONS

### 4.1 Effect of Point Defect Concentration

Intrinsic structural defects can cause severe damage to the elastic properties of the material. Synthesized InSe monolayer might have some inherent defects [47] too. Here, we studied the effect of change of point defect ratio in the sheet by deleting atoms arbitrarily. We have also calculated the error involved in this calculation by performing four different simulations at the same concentration with random atom removal and the possible error that might be present is shown using bar graphs in Fig. 5. We varied the concentration of point defects from 1% to 5% in the structure to observe the change in stress-strain relationship (Fig. 4). The study suggests that as the defect concentration increases, the fracture strength reduces along with the elastic modulus. The reduction of fracture stress, strain and elastic modulus in armchair and zigzag direction for altering defect ratio from 1% to 5% is ~33%, ~16%, ~12%, and ~34%, ~19%, ~19.5%, respectively. Since the vacancy defects always perform as the nucleation points of fracture, occupancy of defects at dispersed and copious locations in the sheet enhances the possibility of creating crack at those points [48]. Thus, the material integrity lessens [49].

.

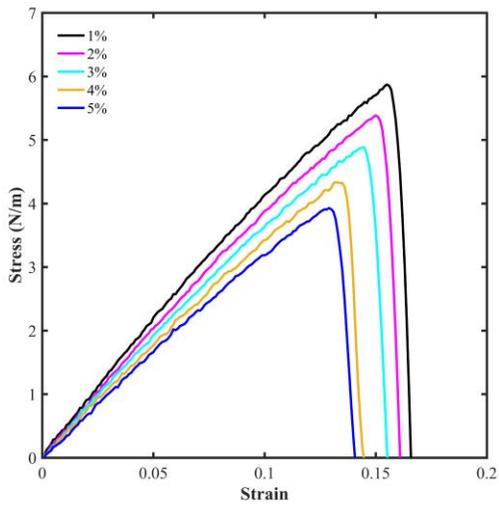 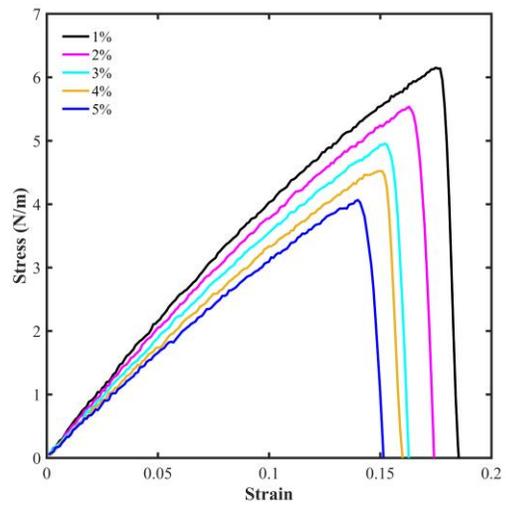

(a) (b)

Fig. 4. Stress-strain relationship for pore concentration when the loading is in (a) Armchair direction (b) Zigzag direction

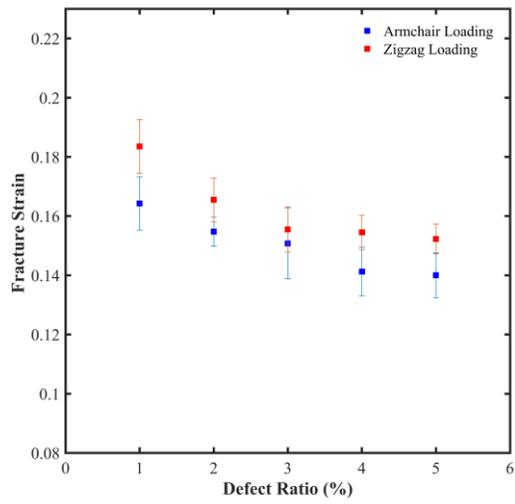 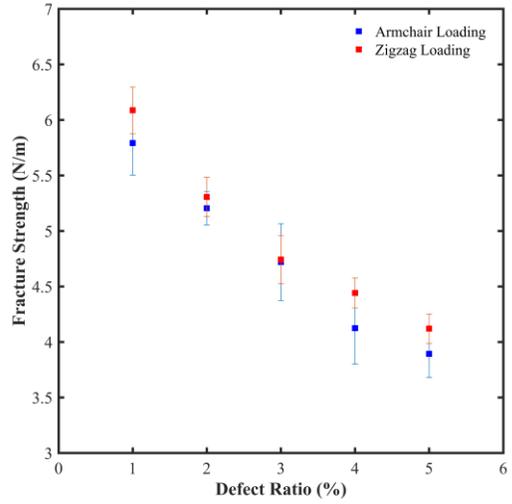

(a) (b)

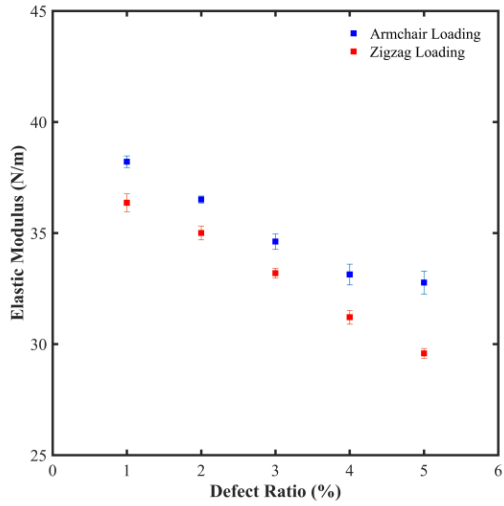

(c)

Fig. 5. Alteration of (a) fracture strain, (b) fracture strength, and (c) elastic modulus of SLInSe with the variation of defect concentration. Bar graphs shows the errors involved in the randomness of atoms deletation.

### 4.2 Stress-strain relationship for different crack lengths:

For pre-cracked sheet, strain is provided in the perpendicular direction of the cracks (armchair tension for zigzag directional crack and vice-versa). Fig. 6 shows the stress-strain relationship of SLInSe with pre-fabricated crack under uniaxial tension at 300 K. It shows that the fracture strength and strain of pre-cracked InSe sheet crucially depends on the crack length. This is because stress localization intensifies around the crack tip as the crack length increases [50]. Thus, material properties such as ultimate stress and failure strain degrades. When pre-fabricated crack is present in the structure the material fails at a lower strain compared to the pristine structure and the fracture strength is also reduced. Although the strength reduces as the pre-crack length increases, the linear regime possesses an almost constant Young's modulus of 43 N/m in armchair directional loading and 42 N/m in zigzag directional loading. As the Young's modulus is almost equal in both the armchair direction and in the zigzag direction, it indicates isotropic behavior [30] of the material but the fracture strength depends on the direction of loading. It can be noted that the armchair directional crack shows more resilience than their zigzag counterpart. It is mainly due to their crack propagation path which has been discussed in details in the fracture mechanism section. Fig. 7 shows the comparison between the armchair and zigzag directional fracture strength and strain.

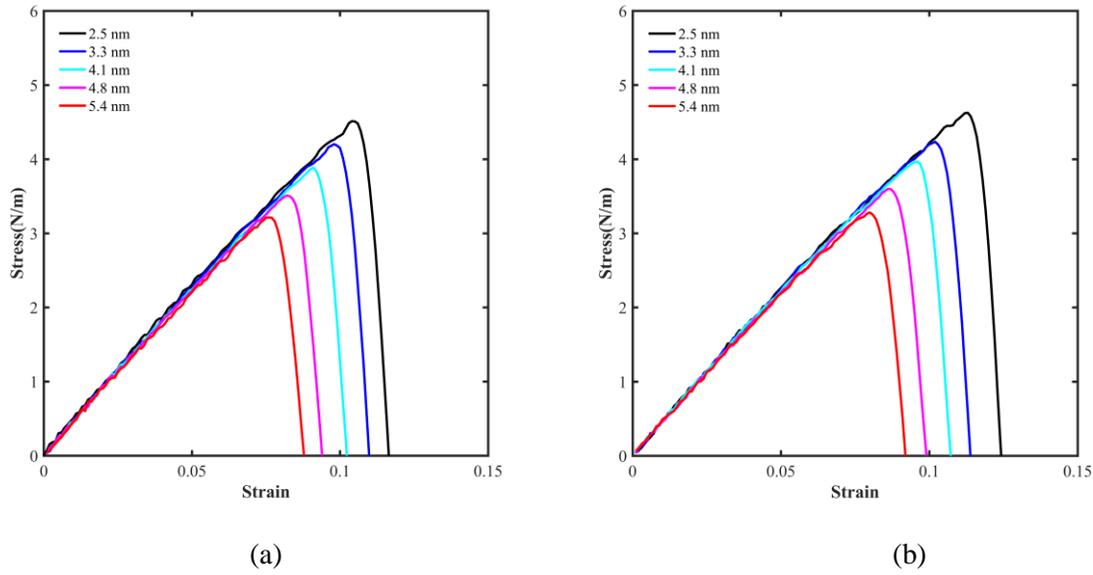

Fig. 6. Stress-strain relationship of SLInSe with initial crack of different lengths when the crack is in (a) Zigzag direction (b) Armchair direction at 300 K

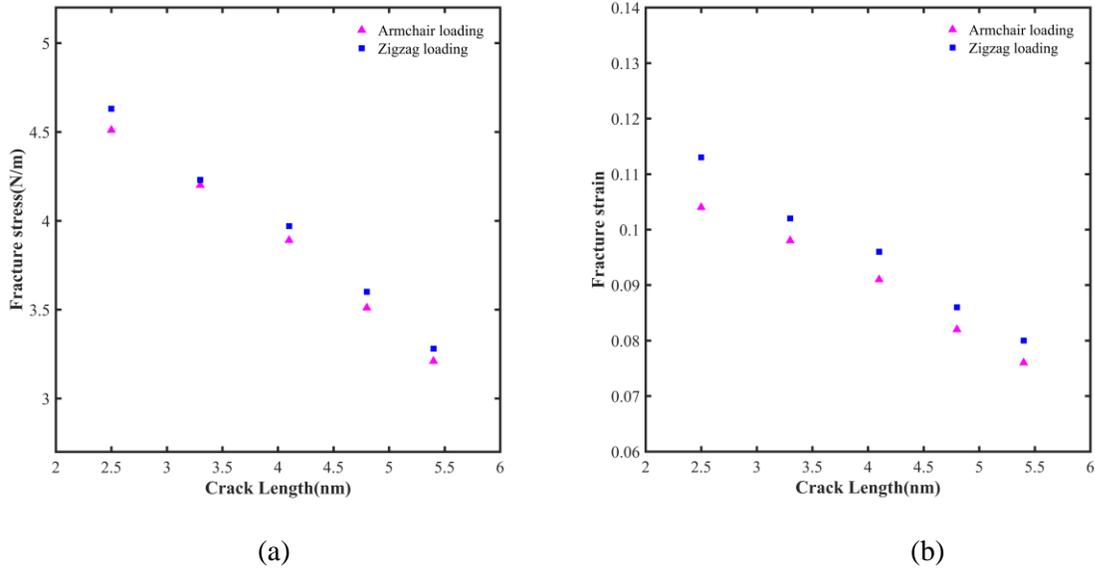

Fig. 7. Change of (a) Fracture stress (b) Fracture strain with initial crack length

### 4.3 Comparison of MD study with Griffith's Theory

Here we considered structures with initial cracks at the center of the monolayer InSe sheet and $a_0/\omega_0 < 0.1$, where $a_0$ denotes initial half crack length and $\omega_0$ denotes initial half sheet length in crack direction, depicted in Fig. 8.

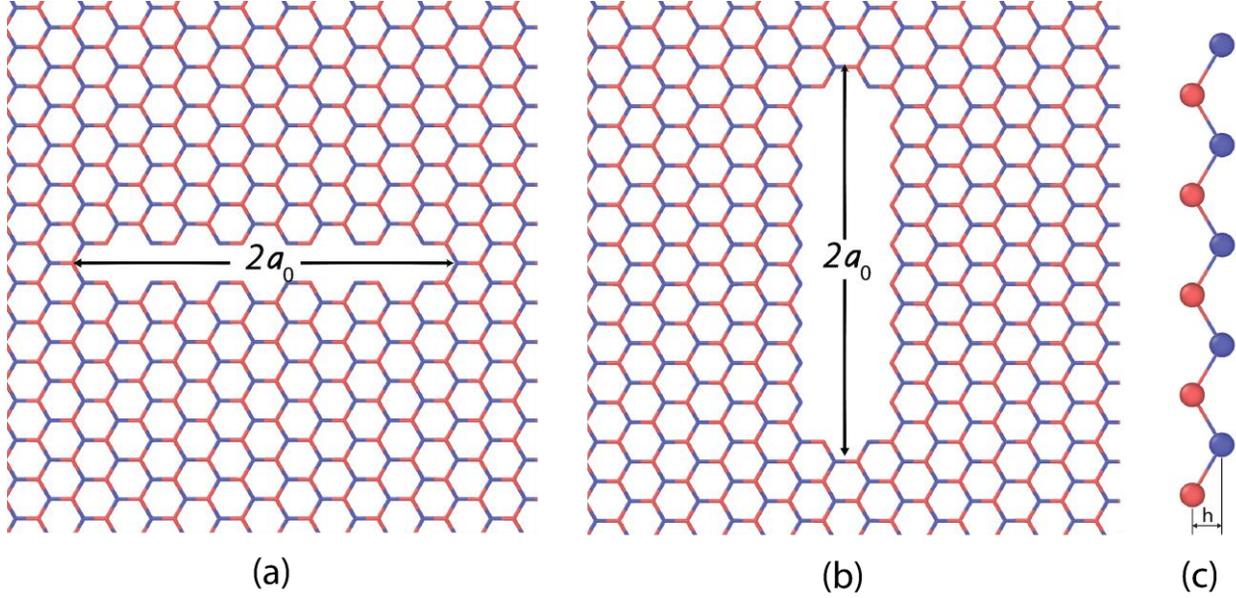

Fig. 8. Top view of pre-cracked SLInSe when the crack is in (a) zigzag direction (b) armchair direction (c) buckling height

We compared our calculated fracture stress from MD with Griffith brittle fracture strength. Fig. 9 indicates the comparison between the evaluated fracture stress from MD and the result from Griffith's Theory and how they vary with the normalized half crack length [51]. From Griffith theory of brittle fracture [50]:

$$\sigma_f = \sqrt{\frac{2\gamma Y}{\pi a_0}} \quad (6)$$

Here, $\sigma_f$ is the fracture stress of the monolayer sheet with pre-crack, $\gamma$ is the surface energy and $Y$ is the Young's modulus calculated from the MD simulation. The surface energy is calculated from the difference of energy between periodic and non-periodic structure in planar direction.

Equation (6) can be rewritten as:

$$\sigma_f \sqrt{\pi a_0} = \sqrt{2\gamma Y} \quad (7)$$

The left side of the equation is termed as fracture toughness and $\sqrt{2\gamma Y}$ is material dependent.

Griffith theory assumes brittle fracture within completely continuum and elastic framework and developed for bulk materials. Nevertheless, InSe has discrete atomic structure, and its bonds are also stretchy. As a result, alteration in edge energy in InSe nano-sheet and flexible nature of bonds eventually result in the variation of fracture stress.

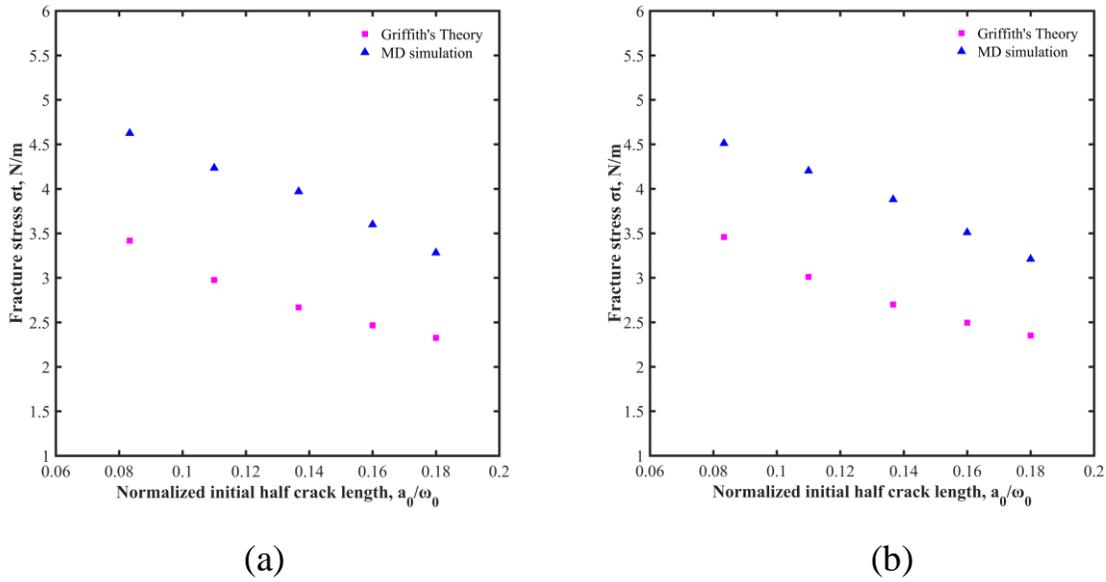

Fig. 9. Variation of fracture stress with the normalized half crack length estimated by Griffith's Theory and MD simulation when the crack is in (a) armchair direction (b) zigzag direction

### 4.4 Stress-strain relationship for different temperature

Temperature is a key factor to drive the structural properties of a material and 2D materials might encounter the threat of temperature during their real-life applications. Thus, it is essential to study the impact of temperature on the mechanical properties of the material. Keeping constant crack length of 4.1 nm, we checked the influence of temperature on the InSe nano-sheet. The stress-strain relationship has been constructed and shown in Fig. 10(a), (b). We varied the temperature by 100 K from 1K to 600K. Increase in temperature deteriorates all the material properties for instance fracture strength, strain, elastic modulus and toughness shown in Fig. 11. The fracture strength, strain, elastic modulus and toughness lowers around ~25%, ~21%, ~5%, ~36% along armchair direction and ~29%, ~26%, ~5%, ~44.5% along zigzag direction for the increase of temperature from 1 K to 600 K accordingly. The change in elastic modulus is almost linear with the changing temperature. Escalation in temperature bolsters higher atomic movements and emboldens the thermal vibrational instabilities. Therefore, the material becomes softer and increased thermal vibration assists some bonds exceeding the critical bond length earlier and initiating pre-mature failure. Furthermore, high temperature is the source of higher entropy in the material which also accelerates the crack propagation. For this reasoning, material strength weakens too[52].

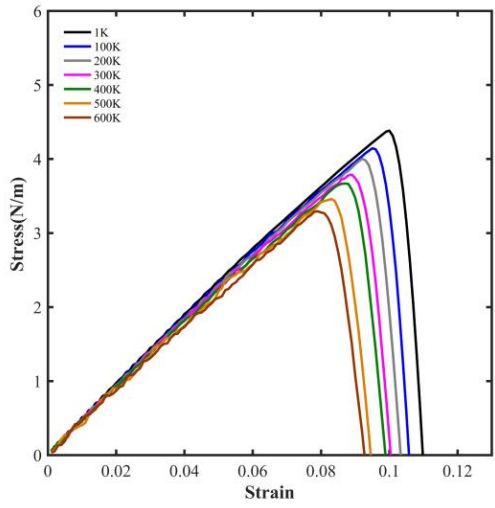 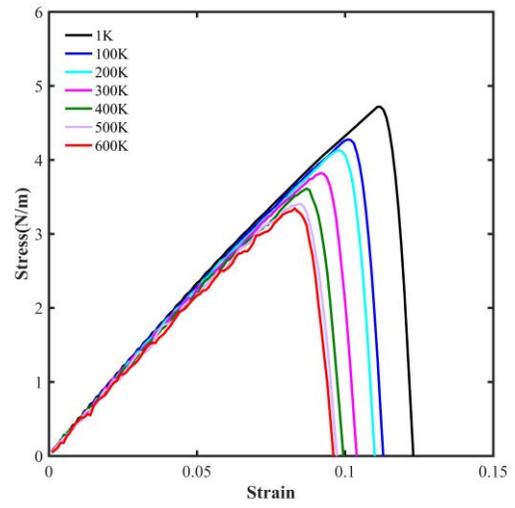

(a)          (b)

Fig. 10. Stress-strain curves of SlInSe showing the temperature dependency when the strain is subjected to (a) armchair direction (b) zigzag direction

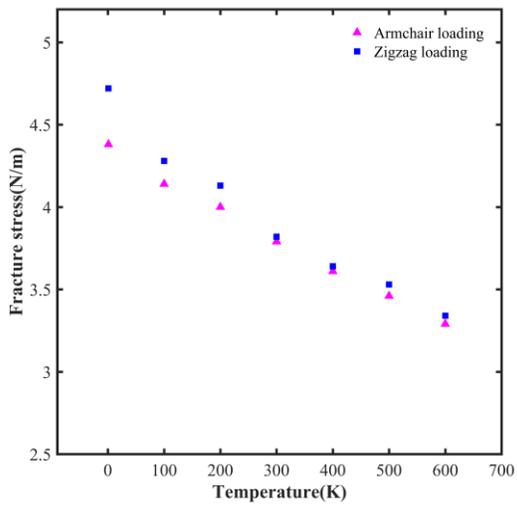 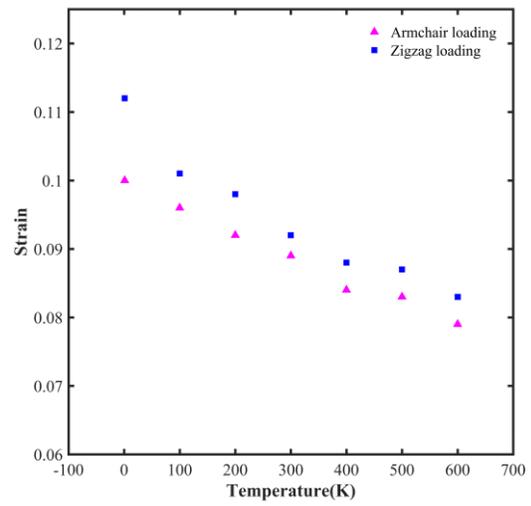

(a)          (b)

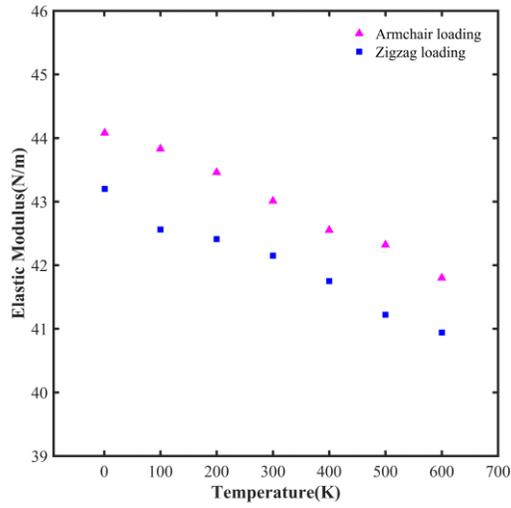
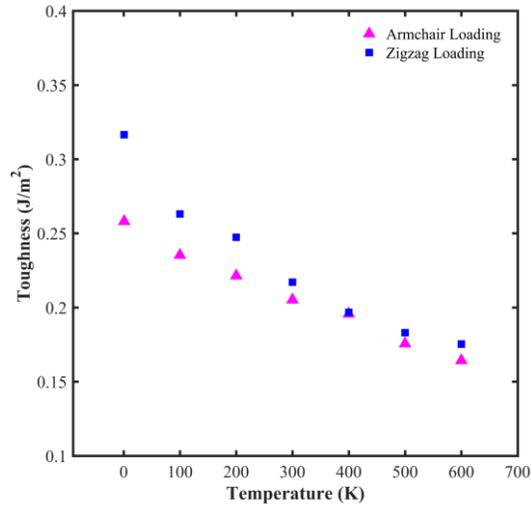

(c)                          (d)

Fig. 11. Temperature dependency of (a) Fracture stress (b) Strain(c) Elastic modulus and (d) Toughness

### 4.5 Fracture mechanism

We investigated the fracture mechanism of defected and pre-cracked samples of InSe monolayer using OVITO visualizing tool. Both of the samples demonstrate almost similar crack propagation scenarios based on the direction of applied strain. The crack propagation mechanism is different for the armchair direction relative to zigzag direction. In the armchair direction four inclined bonds are present symmetrically. When a crack forms after nucleation in case of defected sheet and the crack starts to propagate, these bonds provide two potential crack propagation paths (±60° with x direction). That is why branching phenomenon is observed in case of armchair directional crack propagation [Fig. 12 a(ii), b(ii)]. However, for the crack in zigzag direction, only two bonds are present at the crack tip at a 90° angle. So, when the crack emerges or exits, it only experiences the elongation of the initial one [Fig. 12 a(i), b(i)]. Thus, no branching occurs here in case of zigzag crack[53][54]. Because of this particular feature, the fracture strength in the zigzag direction loading is found to be higher than that in the armchair direction of the monolayer InSe nano-sheet.

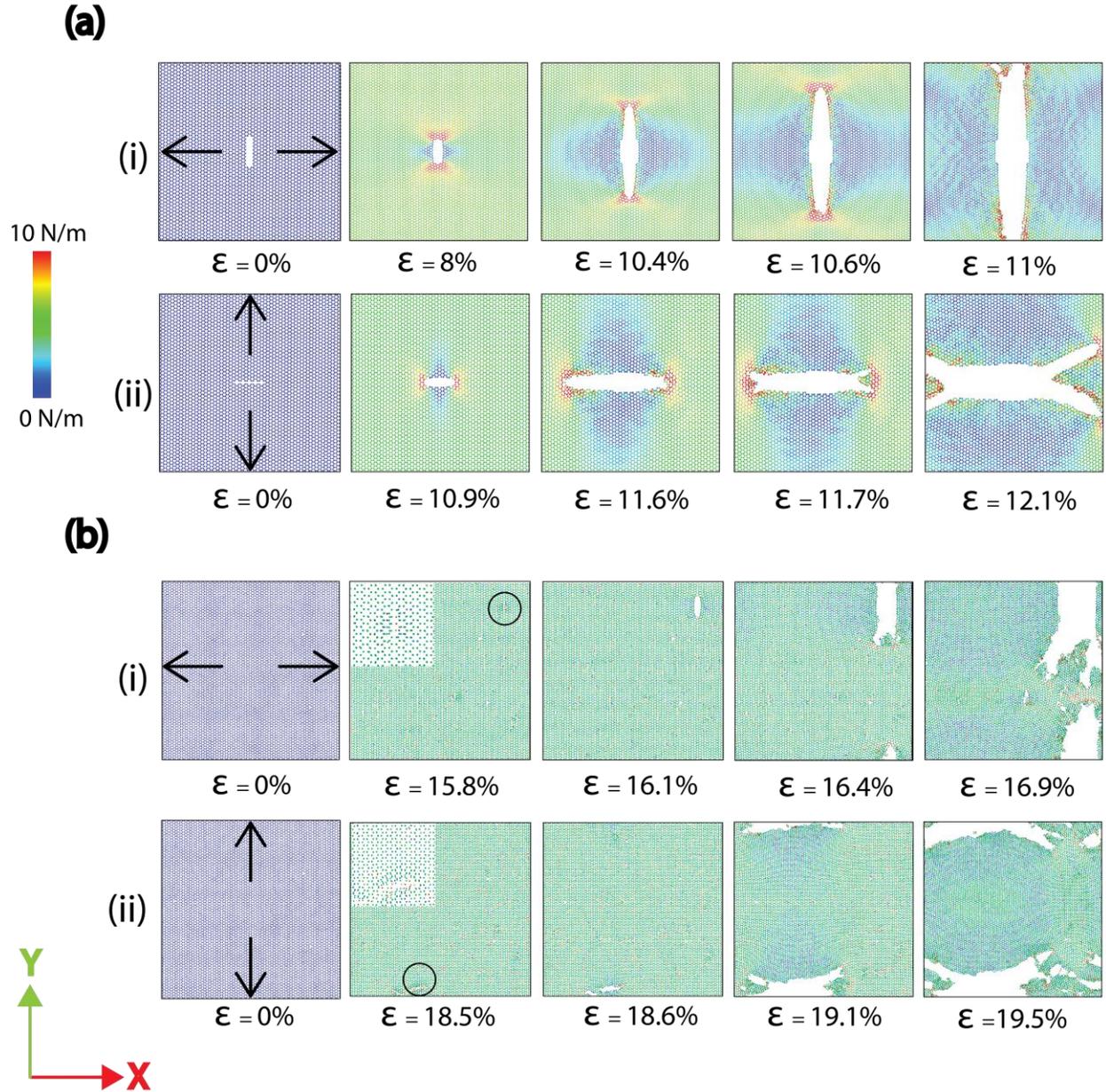

Fig. 12. Crack propagation and distribution of stress in (a) pre-cracked structure (b) point defected structure along (i) armchair (x-axis) (ii) zigzag (y-axis) directional loading. The color bar demonstrates the stress in N/m and the arrows indicate the loading direction

## 5. CONCLUSION

In summary, we employed MD simulations to find the mechanical and the fracture properties of SLInSe sheet with defects. Not only pre-cracks of different lengths are considered but also point vacancy and bi-vacancy defects in the structure are studied. Moreover, we executed simulations at different temperatures keeping a fixed crack in the structure. From the results, it is observed that increasing defect density heavily exacerbates material properties. Lengthening crack in the structure also reduces the fracture properties of the material although elastic modulus remains almost analogous. Nevertheless, Griffith's prediction displays substantial differences from the fracture stress evaluated by MD study, signifying the limitation of the model at the nanoscale. Temperature plays a vital role on the material. Failure stress, strain, elastic modulus, and material toughness all diminishes as the temperature escalates. Fracture mechanism reveals that armchair directional crack can sustain more than that for zigzag directional one. Thus, fracture stress and strain are greater in zigzag directional tension than that of armchair direction as the crack propagate in the perpendicular direction of the loading.

## 6. ACKNOWLEDGEMENT

The authors of this paper would like to thank Multiscale Mechanical Modeling and Research Network (MMMRN) group of BUET for their technical support. We also convey our gratitude to Pritom Bose for his insightful suggestions. M.M.I acknowledges the support from Wayne State University startup funds.

**References:**

[1] K.S. Novoselov, A.K. Geim, S. V. Morozov, D. Jiang, Y. Zhang, S. V. Dubonos, I. V. Grigorieva, A.A. Firsov, Electric field in atomically thin carbon films, Science. (2004). https://doi.org/10.1126/science.1102896.

[2] K.S. Novoselov, A.K. Geim, S. V. Morozov, D. Jiang, M.I. Katsnelson, I. V. Grigorieva, S. V. Dubonos, A.A. Firsov, Two-dimensional gas of massless Dirac fermions in graphene, Nature. (2005). https://doi.org/10.1038/nature04233.

[3] J.C. Meyer, A.K. Geim, M.I. Katsnelson, K.S. Novoselov, T.J. Booth, S. Roth, The structure of suspended graphene sheets, Nature. (2007). https://doi.org/10.1038/nature05545.

[4] A.K. Geim, K.S. Novoselov, The rise of graphene, Nat. Mater. (2007). https://doi.org/10.1038/nmat1849.

[5] H.B. Heersche, P. Jarillo-Herrero, J.B. Oostinga, L.M.K. Vandersypen, A.F. Morpurgo, Manifestations of phase-coherent transport in graphene: The Josephson effect, weak localization, and aperiodic conductance fluctuations, Eur. Phys. J. Spec. Top. (2007). https://doi.org/10.1140/epjst/e2007-00223-7.

[6] H. Oughaddou, H. Enriquez, M.R. Tchalala, H. Yildirim, A.J. Mayne, A. Bendounan, G. Dujardin, M. Ait Ali, A. Kara, Silicene, a promising new 2D material, Prog. Surf. Sci. (2015). https://doi.org/10.1016/j.progsurf.2014.12.003.

[7] M.E. Dávila, L. Xian, S. Cahangirov, A. Rubio, G. Le Lay, Germanene: A novel two-dimensional germanium allotrope akin to graphene and silicene, New J. Phys. (2014). https://doi.org/10.1088/1367-2630/16/9/095002.

[8] Y. Kubota, K. Watanabe, O. Tsuda, T. Taniguchi, Deep ultraviolet light-emitting hexagonal boron nitride synthesized at atmospheric pressure, Science. (2007). https://doi.org/10.1126/science.1144216.

[9] Q.H. Wang, K. Kalantar-Zadeh, A. Kis, J.N. Coleman, M.S. Strano, Electronics and optoelectronics of two-dimensional transition metal dichalcogenides, Nat. Nanotechnol. (2012). https://doi.org/10.1038/nnano.2012.193.

[10] F. Ersan, D. Kecik, V.O. Özçelik, Y. Kadioglu, O.Ü. Aktürk, E. Durgun, E. Aktürk, S. Ciraci, Two-dimensional pnictogens: A review of recent progresses and future research directions, Appl. Phys. Rev. (2019). https://doi.org/10.1063/1.5074087.

[11] H.C. Chang, C.L. Tu, K.I. Lin, J. Pu, T. Takenobu, C.N. Hsiao, C.H. Chen, Synthesis of Large-Area InSe Monolayers by Chemical Vapor Deposition, Small. (2018). https://doi.org/10.1002/smll.201802351.

[12] J. Zhou, J. Shi, Q. Zeng, Y. Chen, L. Niu, F. Liu, T. Yu, K. Suenaga, X. Liu, J. Lin, Z. Liu, InSe monolayer: Synthesis, structure and ultra-high second-harmonic generation, 2D Mater. (2018). https://doi.org/10.1088/2053-1583/aab390.

[13] S. Sucharitakul, N.J. Goble, U.R. Kumar, R. Sankar, Z.A. Bogorad, F.C. Chou, Y.T. Chen, X.P.A. Gao, Intrinsic Electron Mobility Exceeding 103 cm2/(V s) in Multilayer InSe FETs, Nano Lett. (2015). https://doi.org/10.1021/acs.nanolett.5b00493.

[14] A. Politano, G. Chiarello, R. Samnakay, G. Liu, B. Gürbulak, S. Duman, A.A. Balandin, D.W. Boukhvalov, The influence of chemical reactivity of surface defects on ambient-stable InSe-based nanodevices, Nanoscale. (2016). https://doi.org/10.1039/c6nr01262k.

[15] D.A. Bandurin, A. V. Tyurnina, G.L. Yu, A. Mishchenko, V. Zólyomi, S. V. Morozov, R.K. Kumar, R. V. Gorbachev, Z.R. Kudrynskyi, S. Pezzini, Z.D. Kovalyuk, U. Zeitler, K.S. Novoselov, A. Patanè, L. Eaves, I. V. Grigorieva, V.I. Fal'Ko, A.K. Geim, Y. Cao, High electron mobility, quantum Hall effect and anomalous optical response in atomically thin InSe, Nat. Nanotechnol. (2017). https://doi.org/10.1038/nnano.2016.242.

[16] V. Zólyomi, N.D. Drummond, V.I. Fal'Ko, Electrons and phonons in single layers of hexagonal indium chalcogenides from ab initio calculations, Phys. Rev. B - Condens. Matter Mater. Phys. (2014). https://doi.org/10.1103/PhysRevB.89.205416.

[17] P.R. Wallace, Erratum: The band theory of graphite (Phys. Rev. (1947) 71 (622)), Phys. Rev. (1947). https://doi.org/10.1103/PhysRev.72.258.

[18] W. Feng, W. Zheng, W. Cao, P. Hu, Back Gated Multilayer InSe Transistors with Enhanced Carrier Mobilities via the Suppression of Carrier Scattering from a Dielectric Interface, Adv. Mater. (2014). https://doi.org/10.1002/adma.201402427.

[19] S. Lei, F. Wen, B. Li, Q. Wang, Y. Huang, Y. Gong, Y. He, P. Dong, J. Bellah, A. George, L. Ge, J. Lou, N.J. Halas, R. Vajtai, P.M. Ajayan, Optoelectronic memory using two-dimensional materials, Nano Lett. (2015). https://doi.org/10.1021/nl503505f.

[20] S.R. Tamalampudi, Y.Y. Lu, R. Kumar U., R. Sankar, C. Da Liao, K. Moorthy B., C.H. Cheng, F.C. Chou, Y.T. Chen, High performance and bendable few-layered InSe photodetectors with broad spectral response, Nano Lett. (2014). https://doi.org/10.1021/nl500817g.

[21] S. Lei, L. Ge, S. Najmaei, A. George, R. Kappera, J. Lou, M. Chhowalla, H. Yamaguchi, G. Gupta, R. Vajtai, A.D. Mohite, P.M. Ajayan, Evolution of the electronic band structure and efficient photo-detection in atomic layers of InSe, ACS Nano. (2014). https://doi.org/10.1021/nn405036u.

[22] S. Lei, F. Wen, L. Ge, S. Najmaei, A. George, Y. Gong, W. Gao, Z. Jin, B. Li, J. Lou, J. Kono, R. Vajtai, P. Ajayan, N.J. Halas, An atomically layered InSe avalanche photodetector, Nano Lett. (2015). https://doi.org/10.1021/acs.nanolett.5b00016.

[23] W. Li, J. Li, Piezoelectricity in two-dimensional group-III monochalcogenides, Nano Res. (2015). https://doi.org/10.1007/s12274-015-0878-8.

[24] M. Balkanski, C. Julien, M. Jouanne, Electron and phonon aspects in a lithium intercalated InSe cathode, J. Power Sources. (1987). https://doi.org/10.1016/0378-7753(87)80114-3.

[25] C.H. Ho, Y.J. Chu, Bending Photoluminescence and Surface Photovoltaic Effect on Multilayer InSe 2D Microplate Crystals, Adv. Opt. Mater. (2015). https://doi.org/10.1002/adom.201500390.

[26] G.W. Mudd, M.R. Molas, X. Chen, V. Zólyomi, K. Nogajewski, Z.R. Kudrynskyi, Z.D. Kovalyuk, G. Yusa, O. Makarovsky, L. Eaves, M. Potemski, V.I. Fal'Ko, A. Patanè, The direct-to-indirect band gap crossover in two-dimensional van der Waals Indium Selenide crystals, Sci. Rep. (2016). https://doi.org/10.1038/srep39619.

[27] N. Balakrishnan, Z.R. Kudrynskyi, E.F. Smith, M.W. Fay, O. Makarovsky, Z.D. Kovalyuk, L. Eaves, P.H. Beton, A. Patanè, Engineering p–n junctions and bandgap tuning of InSe nanolayers by controlled oxidation, 2D Mater. (2017). https://doi.org/10.1088/2053-1583/aa61e0.

[28] A.R. Goi, A. Cantarero, U. Schwarz, K. Syassen, A. Chevy, Low-temperature exciton absorption in InSe under pressure, Phys. Rev. B. (1992). https://doi.org/10.1103/PhysRevB.45.4221.
[29] C. Song, F. Fan, N. Xuan, S. Huang, G. Zhang, C. Wang, Z. Sun, H. Wu, H. Yan, Largely Tunable Band Structures of Few-Layer InSe by Uniaxial Strain, ACS Appl. Mater. Interfaces. (2018). https://doi.org/10.1021/acsami.7b17247.
[30] X. Chang, H. Li, G. Tang, Tensile mechanical properties and fracture behavior of monolayer InSe under axial tension, Comput. Mater. Sci. (2019). https://doi.org/10.1016/j.commatsci.2018.11.029.
[31] K.J. Xiao, A. Carvalho, A.H. Castro Neto, Defects and oxidation resilience in InSe, Phys. Rev. B. (2017). https://doi.org/10.1103/PhysRevB.96.054112.
[32] Y. Guo, S. Zhou, Y. Bai, J. Zhao, Defects and oxidation of group-III monochalcogenide monolayers, J. Chem. Phys. (2017). https://doi.org/10.1063/1.4993639.
[33] A. Chandran, K.C. George, Defect induced modifications in the optical, dielectric, and transport properties of hydrothermally prepared ZnS nanoparticles and nanorods, J. Nanoparticle Res. (2014). https://doi.org/10.1007/s11051-013-2238-5.
[34] R.A.S.I. Subad, T.S. Akash, P. Bose, M.M. Islam, Engineered defects to modulate fracture strength of single layer MoS2: An atomistic study, Phys. B Condens. Matter. (2020) 412219. https://doi.org/10.1016/j.physb.2020.412219.
[35] MathWorks Announces Release 2018a of the MATLAB and Simulink Product Families, (n.d.). https://www.mathworks.com/company/newsroom/mathworks-announces-release-2018a-of-the-matlab-and-simulink-product-families.html, (n.d.).
[36] T. Hu, J. Zhou, J. Dong, Strain induced new phase and indirect-direct band gap transition of monolayer InSe, Phys. Chem. Chem. Phys. (2017). https://doi.org/10.1039/c7cp03558f.
[37] R. Beardsley, A. V. Akimov, J.D.G. Greener, G.W. Mudd, S. Sandeep, Z.R. Kudrynskyi, Z.D. Kovalyuk, A. Patanè, A.J. Kent, Nanomechanical probing of the layer/substrate interface of an exfoliated InSe sheet on sapphire, Sci. Rep. (2016). https://doi.org/10.1038/srep26970.
[38] P. Hirel, Atomsk: A tool for manipulating and converting atomic data files, Comput. Phys. Commun. (2015). https://doi.org/10.1016/j.cpc.2015.07.012.
[39] M. Ippolito, A. Mattoni, L. Colombo, F. Cleri, Fracture toughness of nanostructured silicon carbide, Appl. Phys. Lett. (2005). https://doi.org/10.1063/1.2081135.
[40] F. Cleri, S.R. Phillpot, D. Wolf, S. Yip, Atomistic Simulations of Materials Fracture and the Link between Atomic and Continuum Length Scales, J. Am. Ceram. Soc. (2005). https://doi.org/10.1111/j.1151-2916.1998.tb02368.x.
[41] A. Stukowski, Visualization and analysis of atomistic simulation data with OVITO-the Open Visualization Tool, Model. Simul. Mater. Sci. Eng. (2010). https://doi.org/10.1088/0965-0393/18/1/015012.
[42] S. Plimpton, Fast parallel algorithms for short-range molecular dynamics, J. Comput. Phys. (1995). https://doi.org/10.1006/jcph.1995.1039.
[43] R. Paul, T. Tasnim, S. Saha, M. Motalab, Atomistic analysis to characterize the impact of temperature and defects on the mechanical properties of germanene sheet, Mater. Res. Express. (2018). https://doi.org/10.1088/2053-1591/aaa73d.
[44] M. Motalab, R.A.S.I. Subad, A. Ahmed, P. Bose, R. Paul, J.C. Suhling, Atomistic Investigation on Mechanical Properties of Sn-Ag-Cu Based Nanocrystalline Solder Material, in: American Society of Mechanical Engineers Digital Collection, 2020. https://doi.org/10.1115/IMECE2019-12109.
[45] S. Mojumder, T. Rakib, M. Motalab, Atomistic study of hardening mechanism in Al-Cu nanostructure, J. Nanoparticle Res. (2019). https://doi.org/10.1007/s11051-019-4530-5.
[46] J.-W. Jiang, Y.-P. Zhou, Parameterization of Stillinger-Weber Potential for Two- Dimensional Atomic Crystals, in: Handb. Stillinger-Weber Potential Parameters Two-Dimens. At. Cryst., 2017. https://doi.org/10.5772/intechopen.71929.

[47] A.A. Kistanov, Y. Cai, K. Zhou, S. V. Dmitriev, Y.W. Zhang, Atomic-scale mechanisms of defect- and light-induced oxidation and degradation of InSe, J. Mater. Chem. C. (2018). https://doi.org/10.1039/c7tc04738j.

[48] A.H.N. Shirazi, R. Abadi, M. Izadifar, N. Alajlan, T. Rabczuk, Mechanical responses of pristine and defective C3N nanosheets studied by molecular dynamics simulations, Comput. Mater. Sci. (2018). https://doi.org/10.1016/j.commatsci.2018.01.058.

[49] Z.D. Sha, Q.X. Pei, Y.Y. Zhang, Y.W. Zhang, Atomic vacancies significantly degrade the mechanical properties of phosphorene, Nanotechnology. (2016). https://doi.org/10.1088/0957-4484/27/31/315704.

[50] Anderson, T. (2017). Fracture Mechanics. Boca Raton: CRC Press, https://doi.org/10.1201/9781315370293, (n.d.).

[51] T. L. Anderson, Fracture Mechanics: Fundamentals and Applications, 3rd edn. (CRC Press, Taylor & Francis, 2005), n.d.

[52] M.A.N. Dewapriya, A. Srikantha Phani, R.K.N.D. Rajapakse, Influence of temperature and free edges on the mechanical properties of graphene, Model. Simul. Mater. Sci. Eng. (2013). https://doi.org/10.1088/0965-0393/21/6/065017.

[53] P. Zhang, L. Ma, F. Fan, Z. Zeng, C. Peng, P.E. Loya, Z. Liu, Y. Gong, J. Zhang, X. Zhang, P.M. Ajayan, T. Zhu, J. Lou, Fracture toughness of graphene, Nat. Commun. (2014). https://doi.org/10.1038/ncomms4782.

[54] M.Q. Le, R.C. Batra, Mode-I stress intensity factor in single layer graphene sheets, Comput. Mater. Sci. (2016). https://doi.org/10.1016/j.commatsci.2016.03.027.